\documentclass[superscriptaddress,showpacs,preprintnumbers,amsmath,amssymb,prl,twocolumn]{revtex4}

\usepackage{graphicx}
\usepackage{dcolumn}
\usepackage{bm}

\begin{document}

\preprint{(Phys. Rev. Lett., accepted)}

\title{Fingerprints for spin-selection rules in the interaction dynamics of O$_2$ at Al(111)}

\author{Christian Carbogno}%
\affiliation{%
Institut f\"ur Theoretische Chemie, Universit\"at Ulm, D-89069 Ulm, Germany
}%
\author{J\"org Behler}%
\affiliation{%
Lehrstuhl f\"ur Theoretische Chemie, Ruhr-Universit\"at Bochum, D-44780 Bochum, Germany
}%

\author{Axel Gro{\ss}}%
\affiliation{%
Institut f\"ur Theoretische Chemie, Universit\"at Ulm, D-89069 Ulm, Germany
}%
\author{Karsten Reuter}%
\affiliation{%
Fritz-Haber-Institut der Max-Planck-Gesellschaft, Faradayweg 4-6, D-14195 Berlin, Germany
}%

\received{18th June 2008}

\begin{abstract}
We performed mixed quantum-classical molecular dynamics simulations based on first-principles potential-energy surfaces to demonstrate that the scattering of a beam of singlet O$_2$ molecules at Al(111) will enable an unambiguous assessment of the role of  spin-selection rules for the adsorption dynamics. At thermal energies we predict a sticking probability that is substantially less than unity, with the repelled molecules exhibiting characteristic kinetic, vibrational and rotational signatures arising from the non-adiabatic spin transition.
\end{abstract}


\maketitle

Chemical interactions are governed by various selection rules, among which the ones due to Wigner formulate the constraint of overall spin-conservation~\cite{wigner27}. For the ubiquitous oxygen gas-phase chemistry a well-known implication of these spin-selection rules is the pronounced inertness of the O$_2$ molecule in its triplet ground state, if the other reactant and the product are spin singlets. Intriguingly, the dissociative adsorption of oxygen at metal surfaces represents exactly such a spin constellation, but there spin-selection rules are rarely perceived as potential reason for a low reactivity. A notable exception is the long-time enigmatic low initial sticking probability of thermal O$_2$ molecules at Al(111) \cite{osterlund97,brune92}, which was recently explained in terms of a corresponding strongly non-adiabatic dissociation dynamics \cite{behler05}: Due to the hindered spin transition impinging O$_2$ molecules maintain their initial triplet configuration and are already repelled back into the gas-phase at rather large distances from the surface where other mechanisms such as charge transfer from the metal are still quite weak. This non-adiabatic picture of the dissociation process was substantiated with extensive locally constrained density-functional theory (DFT) \cite{wu06,behler07} calculations, which allow to localize the triplet spin at the oxygen molecule and thereby give access to the corresponding spin-triplet potential-energy surface (PES). Ensuing molecular dynamics (MD) simulations confining the trajectories of the approaching O$_2$ molecules to this spin-triplet PES revealed indeed a reduced sticking probability $S_{{\rm o}}(E)$ at low kinetic energies $E$, in semi-quantitative agreement with the experimental data and in gross contradiction to the constant and essentially unit sticking probability obtained from equivalent MD simulations on the adiabatic PES.\cite{behler05}

Despite this significant progress to firmly establish the O$_2$ scattering at Al(111) as a first benchmark example for the importance of spin selection rules at metal surfaces, the limitations due to the semi-local DFT exchange-correlation functionals available for the mapping of the corresponding high-dimensional PESs require further attention. At present, we can not rule out that an improved description of electronic exchange and correlation would e.g. not yield barriers on the adiabatic PES and therewith a similarly reduced sticking probability within an entirely adiabatic framework. While this obviously dictates further research along these lines \cite{behler05,mosch08}, one also has to recognize that the hitherto targeted triplet O$_2$ sticking probability is not necessarily an observable that is most sensitive to the non-adiabatic spin flips themselves. In fact, due to the position of the triplet PES barriers at rather large distances from the surface, the shape of $S_{{\rm o}}(E)$ is to a large extent governed by the motion on this PES alone, and it is for this reason that the previous modeling could do even without explicitly accounting for the actual spin transitions themselves.\cite{behler05}

In this Letter we therefore suggest to change the focus to a complementary observable that depends sensitively on the 
spin coupling and which will constitute unequivocal and measurable fingerprints for the existence or non-existence of spin-selection in this system: the initial sticking probability for spin-singlet O$_2$ molecules, $S^*_{{\rm o}}(E)$. As we will demonstrate on the basis of first-principles calculations, this probability is significantly lower than unity at low kinetic energies due to the hindered, but non-zero transitions between the different spin configurations of the impinging O$_2$ molecules. In contrast to the case of the spin-triplet sticking, the repelled molecules leave the surface this time in the flipped spin-state, i.e. they impinge as spin singlets, but leave as spin triplets. The important twist is thus that the repelled molecules have gained the triplet-singlet energy, which in the experiment is about $\Delta E_{\rm TS} \sim 1$\,eV. From the calculated redistribution of this sizable excess energy into translational, rotational and vibrational degrees of freedom we predict that these exiting molecules will therefore lead to signatures that will clearly allow to discriminate them as fingerprints for the non-adiabatic dynamics, even if the molecular beam employed does not entirely consist of singlet O$_2$ molecules. 

Our first-principles calculations of the sticking coefficients are based on the general ``divide and conquer''-style approach and on exactly the same computational setup for the O$_2$/Al(111) system as detailed in Ref. \cite{behler05}. In short, this comprises an extensive mapping of the singlet and triplet PESs along the six molecular degrees of freedom using DFT within the generalized gradient approximation (GGA)~\cite{hammer99}, followed by an interpolation using neural networks ~\cite{lorenz04,behler07b} in order to obtain the continuous representation required for the MD simulations. Locally constrained DFT calculations enforcing a triplet spin at the oxygen molecule are employed to map the triplet PES (corresponding to the $^3\Sigma_g^-$ ground state at large distances from the surface) \cite{behler07}. Due to the improper multiplet representation given by DFT with jellium based exchange-correlation functionals, the same approach but enforcing a singlet spin at the oxygen molecule can not be used to obtain an appropriate description of the lowest-lying singlet PES (corresponding to the $^1\Delta_g$ excited state at large distances from the surface). As discussed before \cite{behler05} an energetically reasonable approximation to this singlet PES is instead obtained through spin-unpolarized DFT calculations, yielding a triplet-singlet energy for the free O$_2$ molecule of $\Delta E_{\rm TS} = 1.2$\,eV. Despite this approximate nature of the singlet state, the different spinors still ensure the orthogonality of the two electronic configurations, which can thus be viewed as spanning a kind of two-dimensional diabatic basis.

\begin{figure}
\includegraphics[clip,width=.95\linewidth]{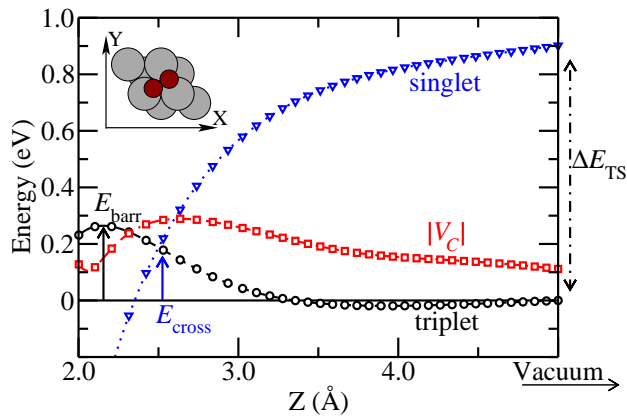}
\caption{\label{fig1}
Calculated singlet- and triplet potential energy along the minimum energy path for a triplet O$_2$ molecule impinging side-on over the Al(111) fcc hollow site, cf. the schematic top view shown as inset. The energy zero corresponds to a free O$_2$ molecule in its triplet ground state. At corresponding distances ${\rm Z} \rightarrow \infty$ far away from the Al(111) surface, the singlet energy is higher by $\Delta E_{\rm TS} \sim 1$\,eV. Additionally shown is the spin coupling matrix element $V_c$ along the path. Obviously, the transitions due to this coupling are strongest at the crossing seam of the singlet and triplet PESs, which is located at rather large distances ${\rm Z} > 2$\,{\AA} away from the surface.}
\end{figure}

Figure \ref{fig1} illustrates the two PESs relevant for the O$_2$ sticking along a one-dimensional cut through configuration space, corresponding to the minimum energy path for a triplet O$_2$ molecule impinging side-on over the Al(111) fcc hollow site. Whereas the preceding work considered only motion confined to one of these electronic states, we now explicitly account for an excited state dynamics and non-adiabatic transitions by performing mixed quantum-classical MD simulations based on the fewest switches algorithm \cite{tully90,carbogno07}. In this surface hopping framework, the nuclear degrees of freedom are integrated classically on one PES at each time step. Simultaneously the density matrix including both electronic states is calculated by integrating the time-dependent Schr\"odinger equation along this trajectory. Transitions from one PES to another are introduced in such a way that for a large number of trajectories the occupation probabilities given by the density matrix are achieved within the smallest number of switches possible. The actual spin coupling matrix element $V_c$ at each point of the configuration space spanned by the six molecular degrees of freedom is calculated by taking advantage of the also available adiabatic spin-polarized ground-state PES \cite{behler05}. Inverting the diagonalization of the diabatic Hamiltonian
\begin{equation*}
\bm{H}=\begin{pmatrix}E_{\rm T} & V_c\\V_c^* & E_{\rm S}\end{pmatrix} \quad ,
\end{equation*}
where $E_{\rm T}$ and $E_{\rm S}$ are the triplet and singlet energies respectively, then uniquely determines the absolute value of the matrix elements $V_c$ in all six considered dimensions, cf. Fig. \ref{fig1}. For the present two-state problem, this thus enables a computationally efficient determination of the non-adiabatic couplings without explicitly requiring to evaluate the underlying wave function dynamics \cite{doltsinis02}. We note, however, that particularly because of the approximate nature of the singlet PES, this procedure likely overestimates the $V_c$ and we will critically discuss this point below.

As a natural first target we use this approach to recompute the sticking curve for impinging triplet O$_2$ molecules that was the central quantity addressed in the preceding study \cite{behler05}. More than 2000 trajectories with random initial molecular orientations were calculated for each kinetic energy, in order to get a reliable statistics for this averaged quantity. As suspected, we obtain a result that is very similar to the curve published in Ref. \cite{behler05}, i.e. a $S_{{\rm o}}(E)$ that is close to zero for thermal molecules, increases monotonically with increasing kinetic energy and approaches 100\% for $E > 0.5$\,eV. While this reconfirms that hindered spin transitions lead indeed to a lowering of $S_{{\rm o}}(E)$ at low kinetic energies as observed experimentally, we also recognize that the explicit consideration of spin transitions in our mixed quantum-classical dynamics barely modifies the result obtained previously with a dynamics that was entirely confined to the triplet PES \cite{behler05}. As apparent from Fig. \ref{fig1}, the lowest-energy molecules are repelled by the barriers on the triplet PES already at distances from the surface, where transitions to the singlet state are energetically forbidden. This determines the characteristic ``S''-shape of the sticking curve, and the finite spin transitions only lead to a small increase of $S_{{\rm o}}(E)$ at intermediate energies between those of the triplet-singlet crossing seam and the triplet barriers, marked as $E_{\rm cross}$ and $E_{\rm barr}$ in Fig. \ref{fig1}. The initial sticking coefficient for triplet O$_2$ molecules at low kinetic energies is thus not very sensitive to the details of the non-adiabatic coupling. This is aggravated by the fact that in this energy window the fraction of repelled molecules, $(1 - S_{{\rm o}}(E))$, consists entirely of O$_2$ molecules in their triplet ground-state. Imagine that the absence of barriers on the hitherto calculated adiabatic PES was an artifact of the employed semi-local DFT exchange-correlation functional, and the real adiabatic PES exhibited barriers just of the same order as the present triplet PES. This would equally yield an ``S''-shaped $S_{{\rm o}}(E)$ curve, explaining the experimental data in an entirely adiabatic framework. Furthermore, since also here the repelled molecules are in their spin-triplet state, it would be impossible to distinguish whether they had been repelled adiabatically or as the result of a hindered spin-transition.

\begin{figure}
\includegraphics[clip,width=.95\linewidth]{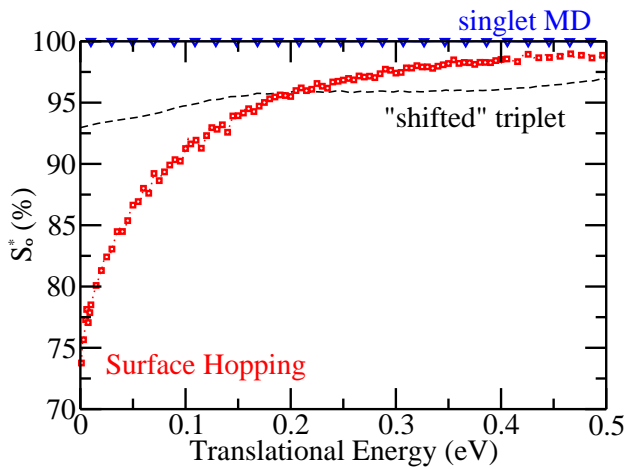}
\caption{\label{fig2}
Initial sticking coefficient at normal incidence for singlet oxygen molecules, $S_{{\rm o}}^*(E)$, as computed for a dynamics confined to the singlet PES ($\triangledown$) and by accounting for non-adiabatic transitions ($\square$). The dashed line represents the sticking coefficient for triplet oxygen, simply shifted by $\Delta E_{\rm TS}$ (see text).}
\end{figure}

This situation is markedly different, if we move on to the sticking curve for a beam of singlet O$_2$ molecules. From the purely attractive character of the corresponding singlet PES, cf. Fig. \ref{fig1}, one would intuitively expect a sticking curve of $S_{{\rm o}}^*(E) \approx 1$ at all kinetic energies, and this is indeed what is obtained when confining the MD simulations to a motion on the singlet PES as shown in Fig. \ref{fig2}. Surprisingly, accounting for the finite singlet-triplet coupling leads now to a significant reduction of the sticking probability at low kinetic energies, cf. Fig.~\ref{fig2}. This is due to the fact that some of the impinging singlet molecules suffer a spin transition near the crossing seam of the singlet and triplet PES, and are back-scattered on the triplet PES, which indeed exhibits some barriers larger than 1\,eV. At increasing kinetic energies, this effect is reduced as more molecules can overcome these effective barriers, but also because the singlet-triplet transition probability is reduced at higher velocities \cite{zener32}.

For the lowest kinetic energies, however, this transition probability is almost 100\%, i.e. virtually all impinging singlet molecules cross to triplet when approaching the singlet-triplet PES crossing seam. One might therefore expect that at these energies the singlet sticking curve simply resembles the triplet sticking curve, just shifted by the vacuum triplet-singlet energy, i.e. $S_{{\rm o}}^*(E) \approx S_{{\rm o}}(E + \Delta E_{\rm TS})$. Again surprisingly, this is not the case and the surface hopping simulations yield an even more pronounced reduction of $S_{{\rm o}}^*(E)$ at the lowest kinetic energies, cf. Fig. \ref{fig2}. This is rationalized by realizing that the coupling matrix elements are not only a function of the molecule-surface separation, but also of the other molecular degrees of freedom. As a consequence, the excess energy released upon transition from singlet to triplet PES is redistributed onto these different degrees of freedom according to the direction of the non-adiabatic coupling vector \cite{tully90}. Correspondingly, not the entire potential energy stored in the electronic singlet configuration is available for the propagation along the reaction path and the dissociation probability is even further reduced. 

\begin{figure}
\includegraphics[clip,width=.95\linewidth]{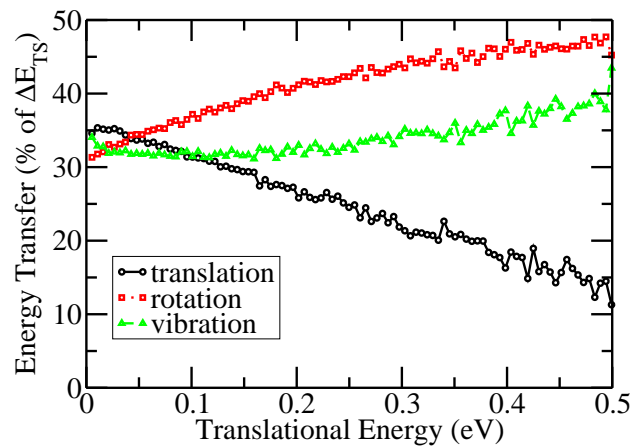}
\caption{\label{fig3}
Percentage of the electronic potential energy $\Delta E_{\rm TS}$ redistributed into translational ($\circ$), rotational ($\square$) and vibrational ($\triangle$) energy of the reflected O$_2$ molecules. Note that all these molecules leave the surface as spin-triplets.}
\end{figure}

A fingerprint for this energy redistribution during the non-adiabatic singlet-triplet transition is given by the amount of energy stored into the translational, vibrational and rotational degrees of freedom of the reflected molecules. Figure \ref{fig3} summarizes this data, and for this it is important to realize that we find essentially all reflected molecules to leave the surface in the triplet spin state. Due to the redistributed excess energy $\Delta E_{\rm TS}$, the reflected
molecules become translationally, vibrationally and rotationally hot, i.e., upon reflection they gain a significant amount of energy in all molecular degrees of freedom. Their energy distribution thus provides unequivocal signatures for the non-adiabatic spin transition. As apparent from Fig. \ref{fig3}, this holds in particular for thermal molecules, for which roughly one third of $\Delta E_{\rm TS}$, thus more than $0.3$ eV, is stored in the translational motion \cite{surfosc}. Even if the experimental molecular beam does not completely consist of singlets, it would therefore easily be possible to discriminate non-adiabatically reflected molecules by their kinetic energies alone.

In contrast to the hitherto much studied sticking of triplet O$_2$ molecules we thus find the scattering of singlet molecules at Al(111) to yield observables in form of the reflected molecules that would clearly flag a potentially non-adiabatic dynamics governed by spin selection rules. If spin transitions were not hindered at all close to the surface, essentially all impinging singlet molecules would stick, since they would quickly release the potential energy stored in the electronic configuration and therewith be able to overcome even modest barriers on the adiabatic PES. With selection rules operating, our calculations predict instead that for example $\sim 20$\% of an impinging beam of thermal singlet molecules will be repelled with characteristic translational, vibrational and rotational signatures. In this respect, we come back to the point that our coupling matrix elements likely overestimate the real transition probabilities, most notably because of the approximate singlet PES and the diabatic Hamiltonian inversion procedure involving the fully adiabatic PES. The resulting too strong destabilization of the singlet-state is particularly pronounced at large molecular distances from the surface due to the spurious charge transfer into the O$_2$ molecule contained in the singlet PES \cite{behler05}. However, even if we reduce the $V_c$ to the value of the vacuum spin-orbit coupling that then yields a gas-phase singlet lifetime that is orders of magnitude longer than the duration of the entire surface scattering event, and consider the absolutely unrealistic case that the $V_c$ stays constant at this minute value even at any closer distance to the surface, we still obtain a reduced, but nevertheless clearly measurable percentage of reflected molecules with distinct kinetic, vibrational and rotational properties \cite{carbogno08}. Moreover, the redistribution of the excess energy into the different degrees of freedom and in particular into the translational motion of the repelled molecules is virtually not changed, so that our conclusions in terms of unequivocal fingerprints remain completely untouched.

Summarizing, we have performed mixed quantum-classical dynamical simulations based on first-principles potential-energy surfaces to demonstrate that the scattering of a beam of singlet O$_2$ molecules at Al(111) will enable an unambiguous assessment of the role of spin-selection rules for the adsorption dynamics. We predict a sticking probability at low kinetic energies that is substantially less than unity, with the repelled molecules exhibiting characteristic kinetic, vibrational and rotational signatures of a non-adiabatic spin transition that will allow to discriminate them even if the molecular beam does not purely consist of spin-singlets. Such experiments are well feasible as recently demonstrated by Burgert {\em et al.} \cite{burgert08}, who already pointed out the importance of spin-selection rules for the interaction of singlet oxygen interacting with small anionic clusters formed of $\sim 10 - 20$ Al atoms. With the here predicted fingerprints, it thus remains to future experimental observation to decide whether spin-selection rules are also operative at the extended metallic surface or not. If yes, the present approach (mixed quantum-classical simulations based on PESs from constrained DFT) sets the conceptual stage to address corresponding effects in the reaction dynamics of more complex systems.

We acknowledge funding from the Deutsche Forschungsgemeinschaft through grants RE 1509/7-1 and GR 1503/17-1.
JB is grateful for financial support by the Academy of Sciences of NRW and by the Fonds der Chemischen Industrie.
We also wish to thank Matthias Scheffler for initiating this project and for his continued interest and support.


\begin{thebibliography}{25}
\bibitem{surfosc}
This value will be reduced when accounting for the recoil of the substrate atoms. On the basis of a three-dimensional surface oscillator model \cite{carbogno07} we estimate a reduction by about a factor of two, which still leaves a clearly measurable translational heating.\cite{carbogno08}
\end{thebibliography}
\end{document}